%% file: main.tex
\documentclass[10pt,letterpaper,table]{article}
\usepackage[top=0.85in,left=2.75in,footskip=0.75in]{geometry}

\usepackage{amsmath,amssymb,gensymb}
\usepackage{IEEEtrantools}

\usepackage{changepage}

\usepackage{textcomp,marvosym}

\usepackage{cite}

\usepackage{nameref}

\usepackage[right]{lineno}

\usepackage{svg}

\usepackage[nopatch=eqnum]{microtype}
\DisableLigatures[f]{encoding = *, family = * }

\usepackage{xcolor}

\usepackage{array}

\newcolumntype{+}{!{\vrule width 2pt}}

\newlength\savedwidth

\usepackage{setspace} 
\usepackage{hyperref}

\raggedright
\usepackage[aboveskip=1pt,labelfont=bf,labelsep=period,justification=raggedright,singlelinecheck=off]{caption}

\makeatletter
\renewcommand{\@biblabel}[1]{\quad#1.}
\makeatother

\usepackage{lastpage,fancyhdr,graphicx}
\usepackage{epstopdf}
\fancyheadoffset[L]{2.25in}
\fancyfootoffset[L]{2.25in}
\renewcommand{\vec}[1]{\boldsymbol{\mathrm{#1}}}
\newcommand{\ped}[1]{_\mathrm{#1}}

\begin{document}
\vspace*{0.2in}

\begin{flushleft}
{\Large
\textbf\newline{Fluid-derived lattices for unbiased modeling of bacterial colony growth} 
}
\newline
\\
Bryan Verhoef\textsuperscript{1,2}*,
Rutger Hermsen\textsuperscript{2,3},
Joost de Graaf\textsuperscript{1,3},
\\
\bigskip
\textbf{1} Institute for Theoretical Physics, Utrecht University, Princetonplein 5, 3584 CC Utrecht, The Netherlands
\\
\textbf{2} Theoretical Biology Group, Department of Biology, Utrecht University, Padualaan 8, 3584 CH Utrecht, The Netherlands
\\
\textbf{3} Centre for Complex Systems Studies, Utrecht University, Leuvenlaan 4, 3584 CE Utrecht, The Netherlands
\bigskip

* b.verhoef@uu.nl

\end{flushleft}
\section*{Abstract}
Bacterial colonies can form a wide variety of shapes and structures based on ambient and internal conditions. To help understand the mechanisms that determine the structure of and the diversity within these colonies, various numerical modeling techniques have been applied. The most commonly used ones are continuum models, agent-based models, and lattice models. Continuum models are usually computationally fast, but disregard information at the level of the individual, which can be crucial to understanding diversity in a colony. Agent-based models resolve local details to a greater level, but are computationally costly. Lattice-based approaches strike a balance between these two limiting cases. However, this is known to come at the price of introducing undesirable artifacts into the structure of the colonies. For instance, square lattices tend to produce square colonies even where an isotropic shape is expected. Here, we aim to overcome these limitations and therefore study lattice-induced orientational symmetry in a class of hybrid numerical methods that combine aspects of lattice-based and continuum descriptions. We characterize these artifacts and show that they can be circumvented through the use of a disordered lattice which derives from an unstructured fluid. The main advantage of this approach is that the lattice itself does not imbue the colony with a preferential directionality. We demonstrate that our implementation enables the study of colony growth involving millions of individuals within hours of computation time on an ordinary desktop computer, while retaining many of the desirable features of agent-based models. Furthermore, our method can be readily adapted for a wide range of applications, opening up new avenues for studying the formation of colonies with diverse shapes and complex internal interactions.

\section*{Author summary}
Bacterial colonies develop highly diverse shapes, ranging from branches to disks and concentric rings. These structures are important because they affect competition between bacteria and evolution in the population. To study the origins and consequences of bacterial colony structures, computational models have been used to great success. However, to speed up simulations, many such models approximate continuous space using regular lattices even though this is known to cause artifacts in the resulting colony shapes. To address this, we explored the use of disordered lattices. We compared two methods from the literature for perturbing a square reference lattice. In some cases, these appeared to work, yet, when the distance between lattice sites, the contact area between cells, and the size of the cells were incorporated into the model, the symmetries of the square reference lattice reappeared. We therefore came up with a method that uses the structure of a dense fluid of disks to generate a disordered lattice. This fluid-derived lattice did not impose undesirable orientational symmetries in any of the models that we tested. Lastly, we show that our approach is very efficient, enabling the simulation of bacterial populations containing millions of individuals on a regular desktop computer.


\section*{Introduction}

In nature, bacteria often form highly structured colonies or biofilms. Colonies grown in laboratory experiments reveal a wide variety of morphologies that depend on environmental conditions, such as substrate hardness and nutrient availability. For example, \textit{Bacillus subtilis} forms disks, branches, concentric rings, and even a chiral branched pattern, where each branch is twisted with the same handedness~\cite{rafols_formation_1998, wakita_self-affinity_1997, ben-jacob_holotransformations_1994}. In some cases, such structures may play a functional role,~\textit{e.g.}, they may affect susceptibility to antimicrobial agents~\cite{brown_resistance_1988, ashby_effect_1994, xu_spatial_1998, ito_increased_2009, frost_cooperation_2018}, prevent invasion by other bacteria~\cite{nadell_extracellular_2015} or increase resistance to predation~\cite{matz_biofilm_2005}. In addition, the development and morphology of colonies has been shown to be important for evolutionary processes. Initially well-mixed populations of \textit{Escherichia coli} form monoclonal patches as they grow out~\cite{hallatschek_genetic_2007}, and the colony morphology affects the genetic diversity that is maintained~\cite{young_lineage_2022}. The position of individuals in the colony and thus the spatial structure of the bacterial population are also important for competition between bacterial strains~\cite{deforet_evolution_2019, zollner_phase_2017} and their ``social'' interactions, including between antibiotic susceptible and resistant strains~\cite{frost_cooperation_2018}. For these reasons, among others, the mechanisms of colony and biofilm growth remain topics of broad interest.

Experiments have provided many valuable insights into these spatially structured processes, but they can be technically challenging to design and control. As a result, numerical modeling has often been used to complement such studies and gain deeper understanding. Three broad approaches to this modeling have emerged: (i)~Early studies into colony morphology formation employed continuous models, where bacteria are represented by a density field~\cite{golding_studies_1998, kozlovsky_lubricating_1999, matsushita_colony_2004, kitsunezaki_interface_1997, schwarcz_uniform_2016}. These models inherently only consider locally averaged and population-level properties of bacteria and specifically discard their individual nature. This is a severe limitation in studying evolutionary processes, where the genotypes and phenotypes of individual organisms play a crucial role. (ii) Individual bacteria are modeled directly in agent-based models. This has the advantage of having full control over internal states and tracking of individuals~\cite{nadell_emergence_2010, lardon_idynomics_2011, bonachela_universality_2011, young_lineage_2022, young_active_2023}. In addition, features such as bacterial shape and mechanical interaction can be described in detail by prescribing the equations of motion~\cite{farrell_mechanically_2013, warren_spatiotemporal_2019, volfson_biomechanical_2008, langeslay_microdomains_2023, los_defect_2022, ghosh_mechanically-driven_2015, grant_role_2014}. However, agent-based modeling is typically computationally costly~\cite{li_nufeb_2019, young_lineage_2022}.
(iii)~An intermediate solution is to constrain the bacteria to move on a lattice~\cite{eden_two-dimensional_1961, ermentrout_cellular_1993, drasdo_coarse_2005, kusch_mollusc_1996, edmonds_mutations_2004, mobius_how_2015, li_specialization_2023, korolev_genetic_2010, kuhr_range_2011, van_dijk_silico_2016, mukhamadiarov_clonal_2024, abrudan_socially_2015}. This sacrifices some details of the local interactions in favor of a substantial speed advantage, as it avoids resolving the full dynamics of the individual bacteria. However, constraining bacteria to a lattice may introduce undesirable features in the solution~\cite{drasdo_coarse_2005, nemati_cellular_2024}, henceforth broadly referred to as `lattice artifacts'.

While these artifacts are often ignored, some earlier research has recognized them~\cite{batchelor_limits_1991, drasdo_coarse_2005} and has addressed them by using various disordered lattices~\cite{moukarzel_vectorizable_1992, drasdo_coarse_2005, tucker_new_2010, schwarcz_uniform_2016}. However, these disordered lattices have in common that they are generated by perturbing a square reference lattice. In addition, potential lattice artifacts have not been quantified in the various models. In this work, we therefore revisit the problem of lattice-based modeling of bacterial colony growth, with the specific goal to simulate a large number of individuals while overcoming lattice-induced symmetry artifacts. Specifically, we explore a class of hybrid lattice-based models~\cite{borer_spatial_2020, gerlee_hybrid_2008, van_hoek_silico_2006, colizzi_evolution_2021}, develop a method to detect and quantify lattice artifacts in the shape of the simulated bacterial colonies using a convex hull, and propose a new way of generating disordered lattices based on the structure of a fluid.

In brief, in this paper we will show that implementing a hybrid lattice-based model on the commonly used square lattice does result in significant lattice artifacts. These cannot be easily eliminated by tuning the neighborhoods of lattice sites. We then compare previously proposed disordered lattices and demonstrate that while these lattices are effective at removing lattice artifacts in models that only take the binary neighborhood relationship between lattice sites into account, they do not fully eliminate them in models that include additional geometric properties of the lattice, such as distances between sites. Additionally, we highlight the challenge of detecting lattice artifacts through visual inspection of the resulting colony shapes alone. Next, we introduce an off-lattice fluid-derived approach to generate disordered lattices and show that, using this method, no lattice-induced symmetries are detectable across the range of models that we have considered. Finally, we demonstrate that hybrid lattice-based models can achieve colony sizes that are difficult to simulate with off-lattice agent-based models. We conclude by discussing how our approach facilitates the simulation of diversification and competition in large-scale colonies.

\section*{Results}

\begin{figure}[!ht]
    \centering
    \includegraphics[width = 1.0\linewidth]{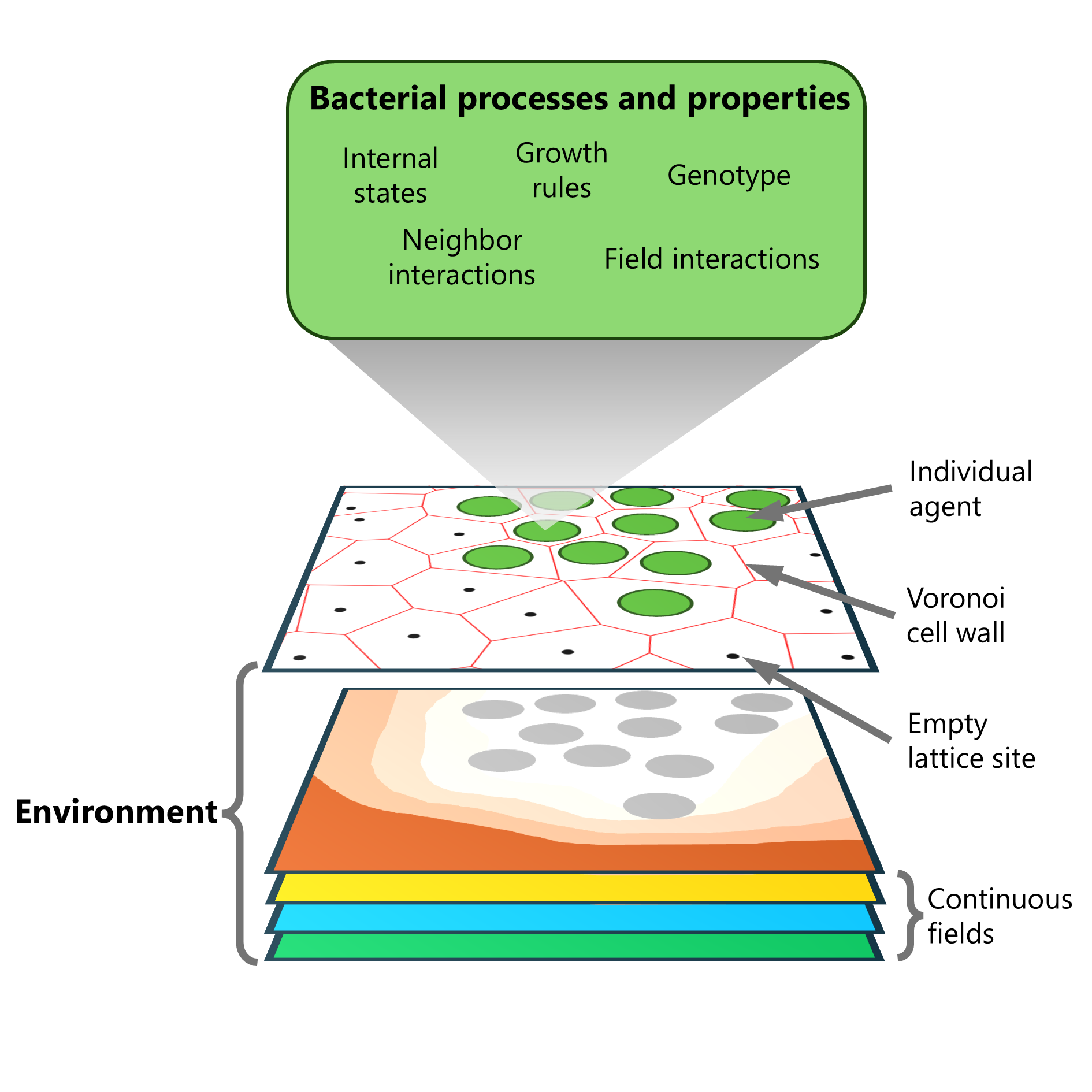}
    \caption{\textbf{Overview of the class of hybrid lattice-based bacterial growth models.} The tilted white field illustrates how bacterial agents (green disks) are constrained to a lattice (black dots). The behavior of the bacteria is determined by various properties and internal processes; the green box lists potential examples. The red lines indicate the Voronoi cells associated with the lattice sites, which define the neighborhood and are used to compute the dynamics of the continuous fields. The bacteria can be coupled to multiple fields, as illustrated by the orange, yellow, blue, and green slices. For example, the orange slice represents the concentration field of a nutrient that is consumed by the bacteria. Its dynamics are governed by a diffusion equation, for which the bacteria serve as sinks, here indicated as gray disks.}
    \label{fig:model_verview}
\end{figure}

We are interested in simulating a class of hybrid lattice-based models, where one or more continuous fields are coupled to discrete bacteria. The bacteria can have a variety of properties and interactions. An overview of this class of hybrid models is given in Fig~\ref{fig:model_verview}. In these hybrid models, a lattice is used to constrain the positions of the bacteria (green disks in Fig~\ref{fig:model_verview}) and to numerically integrate the continuous fields (orange, yellow, blue, and green slices in Fig~\ref{fig:model_verview}). Below, we explore multiple approaches to define lattices and which lattice sites are neighbors. We evaluate these lattices with a minimal model of bacterial colony development, the full details of which are given in the Methods section. This minimal hybrid lattice-based model consists of bacteria coupled to a single continuous nutrient field, governed by a reaction-diffusion equation. The bacteria consume nutrients,~\textit{i.e.}, they acts as sinks to this field. Once a bacterium has consumed an amount of nutrients above a threshold $n_g$, it can divide and produce one daughter cell. This daughter cell is randomly placed on any empty neighboring lattice site. The bacterium is unable to divide when there are no empty neighboring lattice sites available for the daughter cell. During division, the nutrients of the parent cell are split stochastically between parent and daughter.

The remainder of this section is structured as follows. First, we demonstrate that square lattices induce artifacts on resulting colony shapes. We also discuss our method to quantify these lattice artifacts. Next, we show that disordered lattices can diminish the observed lattice artifacts. We investigate two previously proposed methods for generating such a disordered lattice and characterize the quality of the solutions. We notice that there are shortcomings in certain cases, which inspires us to propose a new, fluid-derived method, which we compare in detail with the established methods. We conclude by demonstrating the computational performance of the minimal hybrid model by comparing it directly to an off-lattice (agent-based) approach.

\subsection*{\label{subsec:regular_lattice}Artifacts caused by using square lattices}

\begin{figure}[!ht]
\begin{adjustwidth}{-2.25in}{0in}
    \centering
    \includegraphics{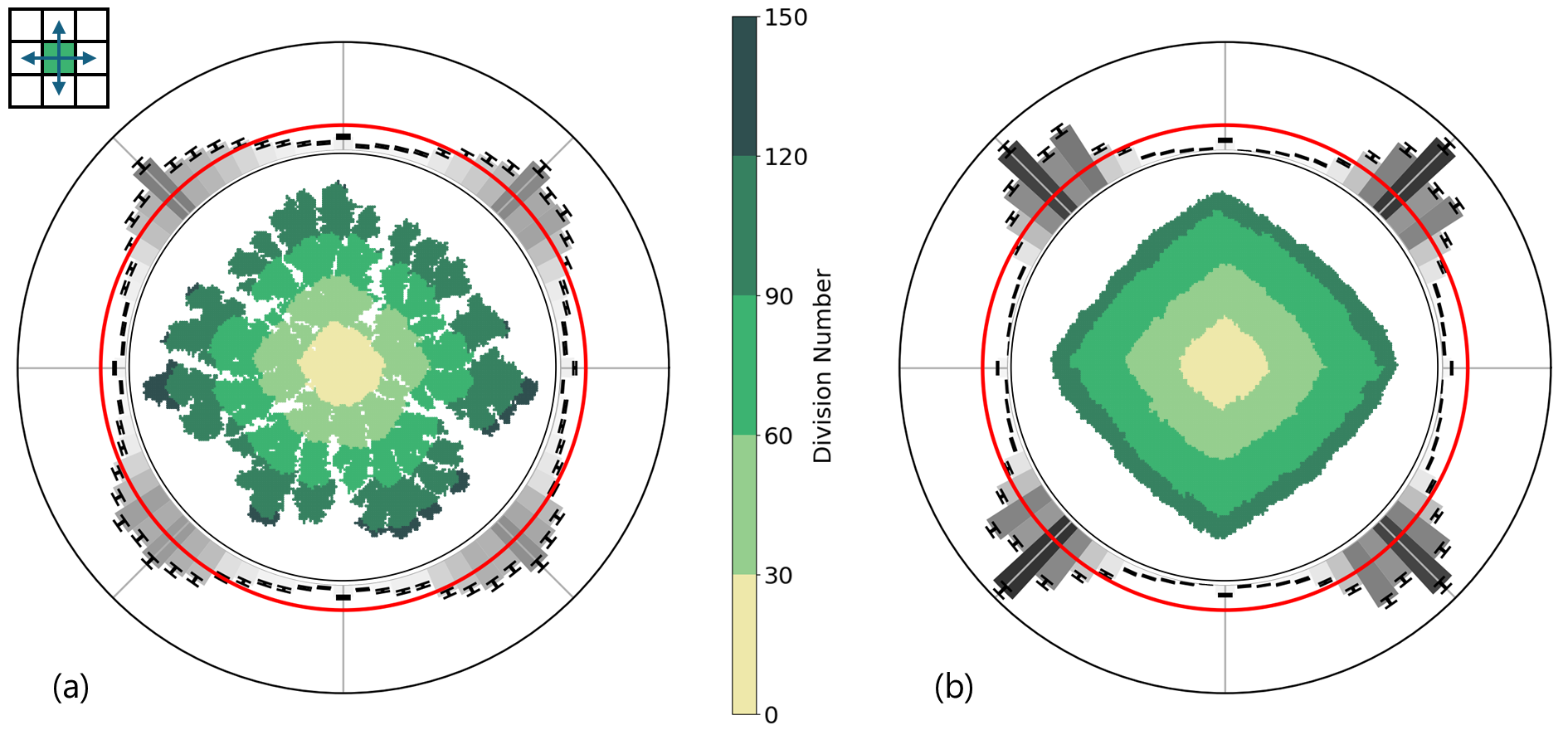}
    \caption{\textbf{Bacterial colony growth according to the minimal hybrid lattice-based model using a square lattice with nearest-neighbor interactions only.} The nearest neighbors on a square grid are as indicated by the blue arrows in the top-left inset. (a) Simulation with low initial nutrient concentration ($c_0 = 0.7 < n_\mathrm{g} / \Delta x^2$) where $n_\mathrm{g}$ is the minimum amount of nutrient that a bacterium needs to consume in order to divide. (b) Simulation with high initial nutrient concentration ($c_0 = 3.0 > n_\mathrm{g} / \Delta x^2$). The remaining parameter choices are specified in Table~\ref{tab:parameters}. The green shapes in the center of both figures show a single representative colony, where the color of each bacterium indicates the number of divisions in the lineage of the bacterium (central color bar). The maximum division number is 131 for colony (a) and 106 for colony (b). The outer ring shows a histogram for the distribution of normal vectors to the convex hulls fitted to colonies. The error bars show the standard error of the mean resulting from 500 independent realizations. The gray scale serves to guide the eye to regions of high incidence. The red circle provides the ideal, isotropic, distribution with value $(2\pi)^{-1}$. In both figures, the histograms indicate that the colonies tend to have a diamond shape.}
    \label{fig:grow_square}
\end{adjustwidth}
\end{figure}

We started by performing simulations of the hybrid model on a square lattice, where we only include nearest neighbors. This choice is also known as the Von Neumann neighborhood and is shown in the inset of Fig~\ref{fig:grow_square}. Two distinct environmental conditions were considered by changing the initial (reduced) nutrient concentration $c_0$: a nutrient-poor environment with $c_0 = 0.7$ and a nutrient-rich environment with $c_0 = 3.0$. The simulation units and remaining parameters can be found in Table~\ref{tab:parameters}.

\begin{table}[!ht]
\begin{adjustwidth}{-2.25in}{0in}
    \centering
    \caption{\textbf{Parameters and constants used in the minimal hybrid lattice-based model.} Default values and their simulation units, and an estimate of simulation values in physical units are given. The maximum self consumption time is used to link the simulation time unit $\Delta t$ to the physical unit.}
    \begin{tabular}{l l l l}
        \hline
         Parameter & Simulation value & In physical units & Literature value \\
         \hline
         Max self consumption time & 25 ($\Delta t$) & 25 min & 8.3 min \cite{schwarcz_uniform_2016}\\  
         $n_\mathrm{g}$ & 1 ($n_\mathrm{g}$) & 12 pg & 36 pg \cite{schwarcz_uniform_2016}\\
         $v_\mathrm{max}$ & 0.04 ($n_\mathrm{g} / \Delta t$) & 0.48 pg/min & --\\
         $K$ & 1 ($n_\mathrm{g} / \Delta x^2$) & 0.74 g/L & --\\
         $c_0$ & 0.5 - 3.0 ($n_\mathrm{g} / \Delta x^2$) & 0.52 - 2.22 g/L & 0.5 - 10 g/L \cite{wakita_self-affinity_1997}\\
         $D_c$ & 0.04 ($\Delta x^2 / \Delta t$) & $6\cdot 10^{-12}$ cm$^2$/s & $3\cdot 10^{-12}$ cm$^2$/s \cite{rana_spreading_2017} \\
         \hline
    \end{tabular}  
    \label{tab:parameters}
\end{adjustwidth}
\end{table}

The results of these simulations are given in Fig~\ref{fig:grow_square}. We can distinguish two colony morphologies: branched in a nutrient-poor environment (Fig~\ref{fig:grow_square}a) and dense in a nutrient-rich environment (Fig~\ref{fig:grow_square}b). These morphologies look qualitatively similar to the morphologies observed for \textit{B. subtilis} colonies~\cite{rafols_formation_1998, kawasaki_modeling_1997}, and in other numerical studies~\cite{golding_studies_1998, matsushita_formation_1999, nadell_emergence_2010, schwarcz_uniform_2016, rana_spreading_2017}. 
However, it is visually clear that both colonies possess an overall diamond shape. This diamond shape is clearly an artifact from the underlying lattice, in line with previous work on this topic~\cite{freche_surface_1985, batchelor_limits_1991, drasdo_coarse_2005}.

To quantify the discrete orientational symmetries in Fig~\ref{fig:grow_square}, we repeated these simulations 500 times with random seeds and fitted a convex hull to each of the resulting colonies. We then determined the normal vector to each segment of the convex hulls. Subsequently, we created a histogram of the angles of these vectors, where each vector was weighted by the length of the associated convex hull segment. This histogram is shown in the outer ring of the panels of Fig~\ref{fig:grow_square}. A more detailed description of this procedure is given in the Methods section along with an overview in~\nameref{S2_Figure}. These distributions show that the shapes of the colonies in both the nutrient-poor and nutrient-rich environments are significantly different from the isotropic situation, which is indicated by the red circle. Furthermore, the previously observed four-fold symmetry of the diamond shape is clearly exposed.

We attempted to diminish these lattice artifacts by changing the definition of the neighborhood of a lattice site. Instead of the Von Neumann neighborhood, we now use the Moore neighborhood, which includes both the four nearest neighbors and the four next-nearest (diagonal) neighbors. To tune the relative frequency with which daughter cells are placed on a next-nearest neighbor site, we assign a weight $w\ped{d}$ to the next-nearest neighbor. Whenever division occurs, an empty site is then selected from the ones neighboring the parent cell, with the probability of the next-nearest neighbor sites weighted by $w\ped{d}$. The value of $w\ped{d}$ affects the shape of the resulting colony and can be tuned to reduce the lattice artifacts.

\begin{figure}[!ht]
\begin{adjustwidth}{-2.25in}{0in}
    \centering
    \includegraphics[width=1.0\linewidth]{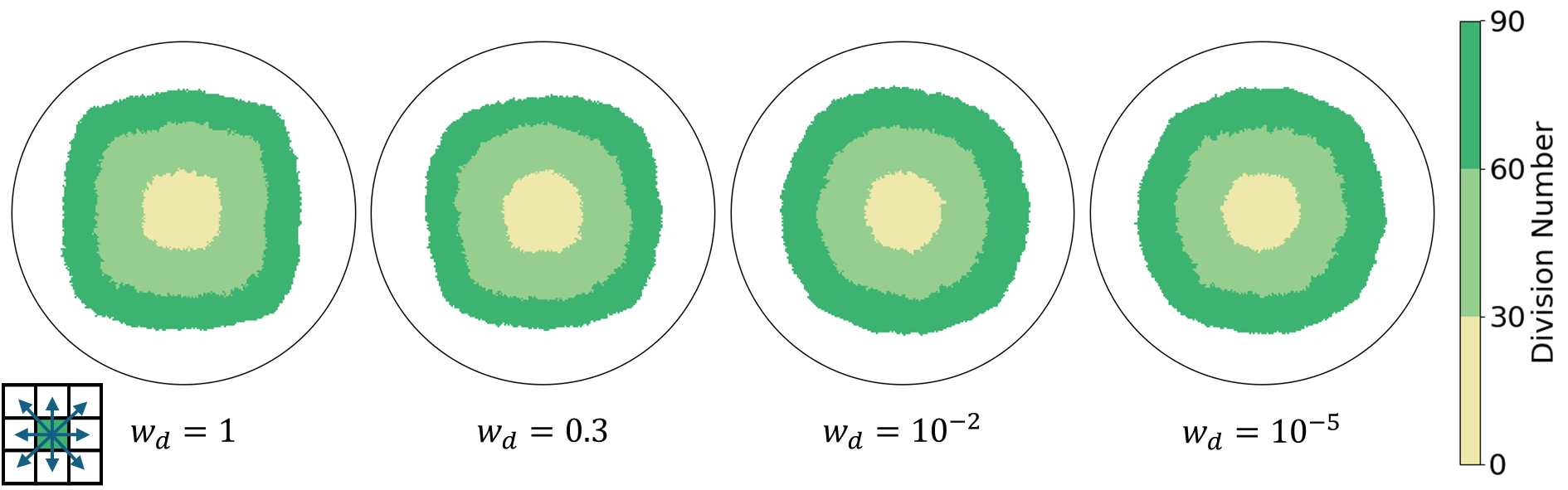}
    \caption{\textbf{Bacterial colony growth on a square lattice with various next nearest-neighbor interaction weights.} Simulations were all performed at high initial nutrient concentration, as in Fig~\ref{fig:grow_square}b. The neighborhood of a lattice site includes nearest and next-nearest neighbors as indicated in the bottom-left inset. The probability that a daughter cell is placed on a next nearest-neighbor site is weighted by $w\ped{d}$, where $w\ped{d} = 1$ gives equal probability between placement on next-nearest and nearest-neighbor sites. The representation and parameter choices are otherwise the same as in Fig~\ref{fig:grow_square}.}
    \label{fig:diag_weighted}
\end{adjustwidth}
\end{figure}

In Fig~\ref{fig:diag_weighted}, a selection of dense colonies with high initial nutrient concentration ($c_0 =3$) simulated with various choices of $w\ped{d}$ is given. It should be noted that we only apply the Moore neighborhood to the growth algorithm. The reaction-diffusion equations are solved on the Von Neumann neighborhood as before. For $w\ped{d} = 1$, we obtain square colonies instead of the diamond-shaped colonies of Fig~\ref{fig:grow_square}. The resulting colonies become less square and more circular as we decrease $w\ped{d}$. Perhaps surprisingly, we do not recover the diamond shape when we lower $w\ped{d}$. The reason is that, even if $w\ped{d}$ is infinitesimally small, growth in diagonal directions still occurs when none of the lateral sites is empty. The diamond shape appears when there is no growth in the diagonal direction at all.

\begin{figure}[!ht]
\begin{adjustwidth}{-2.25in}{0in}
    \centering
    \includegraphics{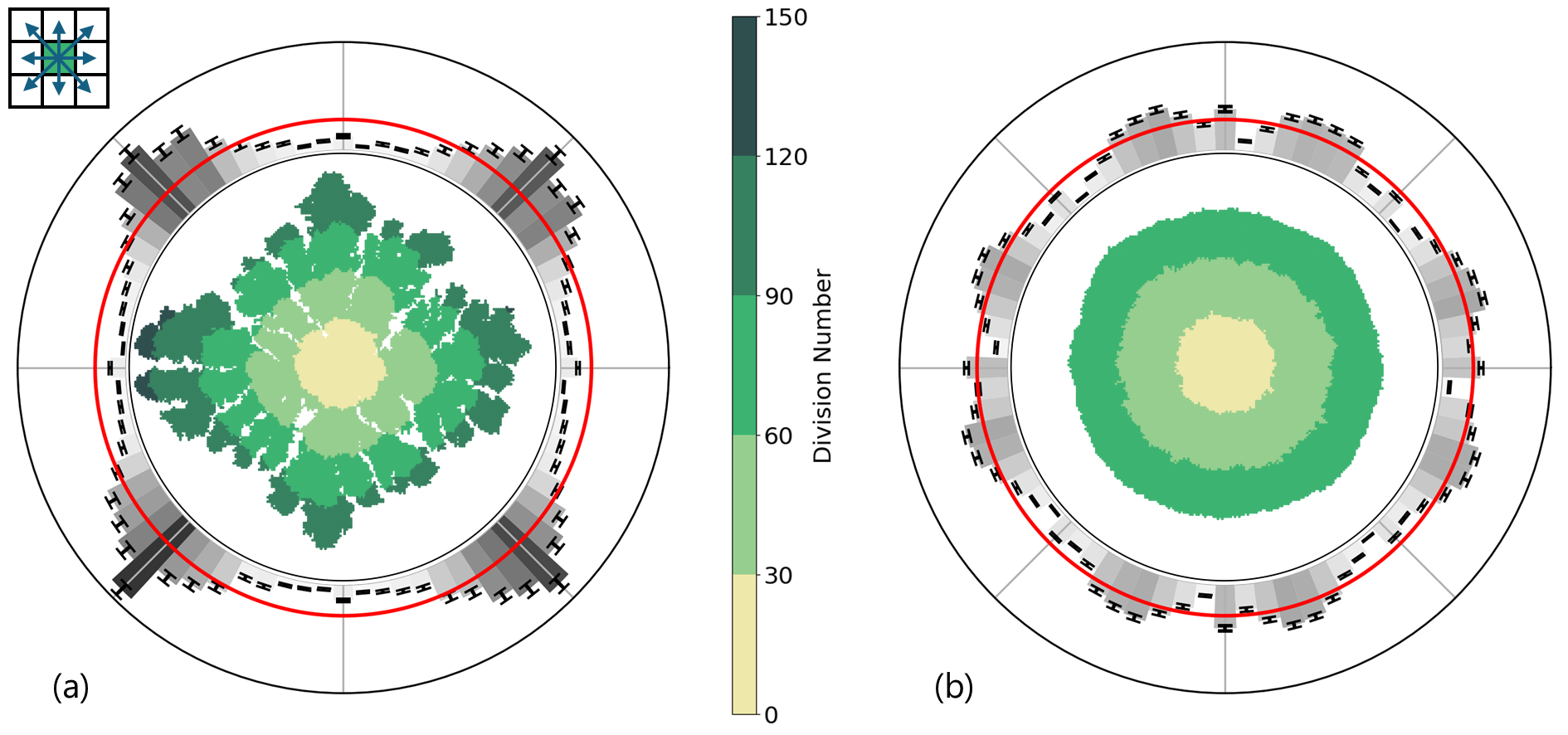}
    \caption{\textbf{Bacterial colony growth on a square lattice with next nearest-neighbor interactions.} The neighborhood of a lattice site is indicated by the blue arrows in the top-left inset. The weight of growth into the diagonal next nearest-neighbor sites was set to $w\ped{d} = 10^{-5}$. The selection of parameters and representation is otherwise the same as in Fig~\ref{fig:grow_square}. (a) For the low initial nutrient concentration the maximum division number is 129. (b) The maximum division number is 86 for the high initial nutrient concentration. In both figures, the histograms demonstrate that the shapes of the colonies are affected by the symmetries of the lattice.}
    \label{fig:grow_square_diag}
\end{adjustwidth}
\end{figure}

To assess whether lattice artifacts are resolved with this method we applied the convex hull procedure to the $w\ped{d} = 10^{-5}$ case, as this seemed to provide the most circular colonies. We considered both the nutrient-poor and nutrient-rich environments. The results are shown in Fig~\ref{fig:grow_square_diag}. In the nutrient-poor environment of Fig~\ref{fig:grow_square_diag}a, the diamond-shaped lattice artifacts we observed in Fig~\ref{fig:grow_square} remain. The numerical solution of the nutrient diffusion equation contains the same lattice artifacts and seems to induce them in the final colony shape. Furthermore, in the nutrient-rich environment of Fig~\ref{fig:grow_square_diag}b, the diamond shape is no longer visible. However, the histogram around the dense morphology in Fig~\ref{fig:grow_square_diag}b still shows clear discrete orientational symmetry.  Even when growth is not constrained by nutrients, we were not able to tune the model to remove the orientational symmetry imposed by the lattice.

\newpage
Taken together, these findings suggest that modifying the growth algorithm alone is not sufficient to remove symmetry-imposing lattice artifacts. Instead of changing the model, a better approach might be to change the underlying lattice. While one could consider other regular lattices, the symmetries of the triangular lattice used by Kozlovsky~\textit{et al.}~\cite{kozlovsky_lubricating_1999} in their continuous model for colony growth remain evident in some of their solutions. Therefore, we expect that other regular lattices should produce similar artifacts.

\subsection*{\label{sec:disordered_lattice}Disordered lattices remove discrete orientational symmetries in colony growth}

The artifacts that we observed are caused by the the underlying symmetries of the lattice. We used a square lattice, which has four-fold orientational order, which is expressed in the shape of the resulting colonies. A possible solution to overcome these artifacts is to use a disordered lattice. Ideally, such a lattice should obey certain requirements. In particular, the lattice should have a minimum distance between sites. This ensures that bacteria have a minimum size, and also prevents that the integration of the reaction-diffusion equations become unstable~\cite{chan_stability_1984, duchemin_explicitimplicitnull_2014}. Other than that, the lattice should not possess additional symmetries or long-range order. Creating a disordered lattice with the desired properties requires some care. We start by detailing the steps involved in generating the lattice before turning to the main result of this work.

\begin{figure}[!ht]
\begin{adjustwidth}{-2.25in}{0in}
    \centering
    \includegraphics[scale=1.15]{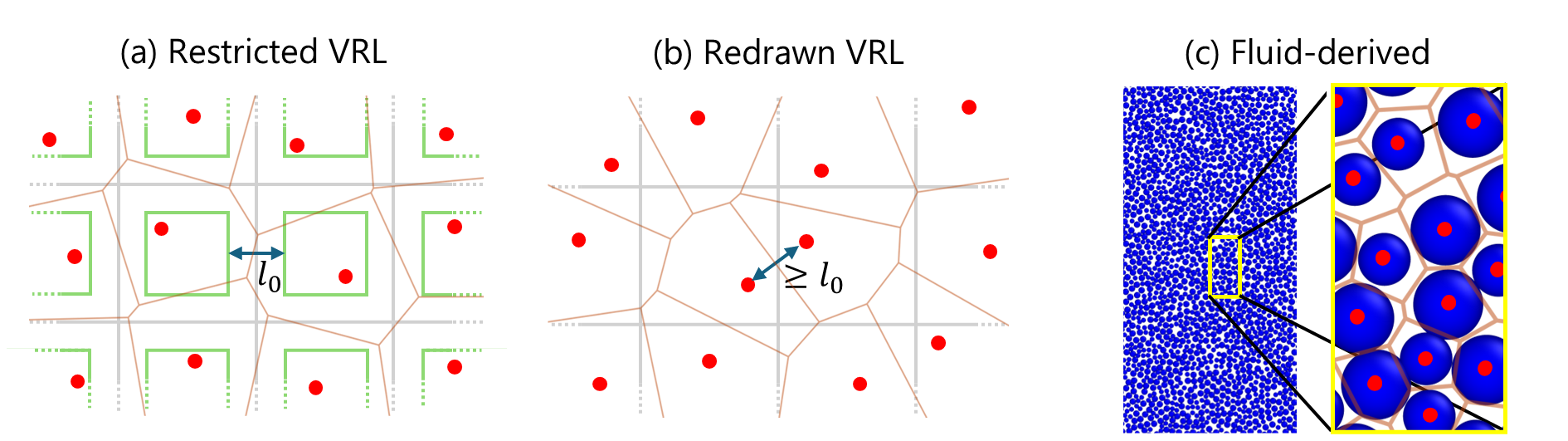}
    \caption{\textbf{Three methods of generating pseudo-random coordinates for a disordered lattice.} The red dots indicate the randomly drawn lattice sites and the orange lines indicate the associated Voronoi tessellation in all figures. For the (a) restricted and (b) redrawn Vectorizable Random Lattice (VRL) proposed by Moukarzel and Herrmann~\cite{moukarzel_vectorizable_1992}, and Tucker~\cite{tucker_new_2010} respectively, the gray squares indicate the reference cells. Each reference cell contains exactly one lattice site. The blue double arrow indicates the tuneable minimum separation between lattice sites $l_0$. In (a) the minimal separation $l_0$ is ensured by randomly placing the lattice sites inside the subdomains of the reference cell indicated as green squares. Our new method (c) uses snapshots of an off-lattice fluid of bidisperse soft disks to generate the disordered grid.}
    \label{fig:lattice_gen}
\end{adjustwidth}
\end{figure}

Several methods have been proposed for generating a set of pseudo-random coordinates to construct a disordered lattice (Fig~\ref{fig:lattice_gen}). The Poisson random lattice proposed by Christ~\textit{et al.}~\cite{christ_random_1982} consists of a set of randomly chosen points with uniform distribution. A limitation of this method is that there is no minimum distance between neighbors. Without such a minimum distance, the effective size of bacteria determined by the lattice varies significantly. In addition, the solution to the continuous part of the hybrid model can become numerically unstable. Moukarzel and Herrmann~\cite{moukarzel_vectorizable_1992} improved on this with their vectorizable random lattice (VRL). A schematic overview of this lattice is given in Fig~\ref{fig:lattice_gen}a. The pseudo-random coordinates of the lattice are generated by placing one site of the disordered lattice randomly in each of the cells of a square reference lattice. A minimum distance between sites $l_0$ can be imposed by restricting the placement domains to a subsquare centered in each cell. We will refer to this lattice as the restricted VRL. Tucker~\cite{tucker_new_2010} adapted the VRL by changing the way $l_0$ is realized. Instead of restricting where a site can be placed within each reference cell, the positions of two sites are redrawn when they are too close together. This lattice will be referred to as the redrawn VRL and is shown in Fig~\ref{fig:lattice_gen}b.

Another system that is known to be disordered and have a well behaved particle-particle separation is a dense fluid of disks~\cite{tanaka_revealing_2019}. Therefore, we turned to a standard model fluid to generate the coordinates to our grid. We simulated a dense two-dimensional (2D) system of bidisperse Weeks-Chandler-Anderson (WCA) particles~\cite{weeks_role_1971}. We chose a bidisperse system to prevent crystallization at high densities~\cite{watanabe_size-dispersity_2005}. After equilibration, we took a snapshot of the configuration and used the particle centers of mass as sites in our disordered lattice. An overview of this fluid-derived lattice is given in Fig~\ref{fig:lattice_gen}c and a more detailed description of the fluid simulations can be found in the Methods section.

For all three disordered lattices, we generated a regular Voronoi tessellation from the generated pseudo-random coordinates. This partitioned our 2D domain into cells. Two lattice sites were considered neighbors when their associated Voronoi cells share a wall.

We computed the probability density functions (PDF) of distances between neighboring lattice sites and of Voronoi cell volumes for all three types of disordered lattices to check whether they meet our requirements. Here, we define our unit of distance $\Delta x$ as the average distance between lattice sites. In addition, we investigated isotropy of the lattices by computing the pair distribution function 
\begin{align}
    \rho^{(2)}(\vec{x}) = \left\langle \sum_{i=1}^N \sum_{j\neq i}^N \delta(\vec{x}_i) \delta(\vec{x} - \vec{x}_j) \right\rangle,
\end{align}
where $\vec{x}_i$ is the position of lattice site $i$ and $N$ the total number of lattice sites. Expressed in polar coordinates, $\rho^{(2)} (r, \theta)$ is a measure of the probability there is a site at a distance $r$ and angle $\theta$ from another site. For isotropic lattices $\rho^{(2)} (r,\theta) = \rho^2 g(r)$, where $\rho = N/A$ with $A$ the total area of the lattice and $g(r)$ the pair correlation function commonly used to describe fluids.

\begin{figure}[!ht]
\begin{adjustwidth}{-2.25in}{0in}
    \centering
    \includegraphics[scale = 1.15]{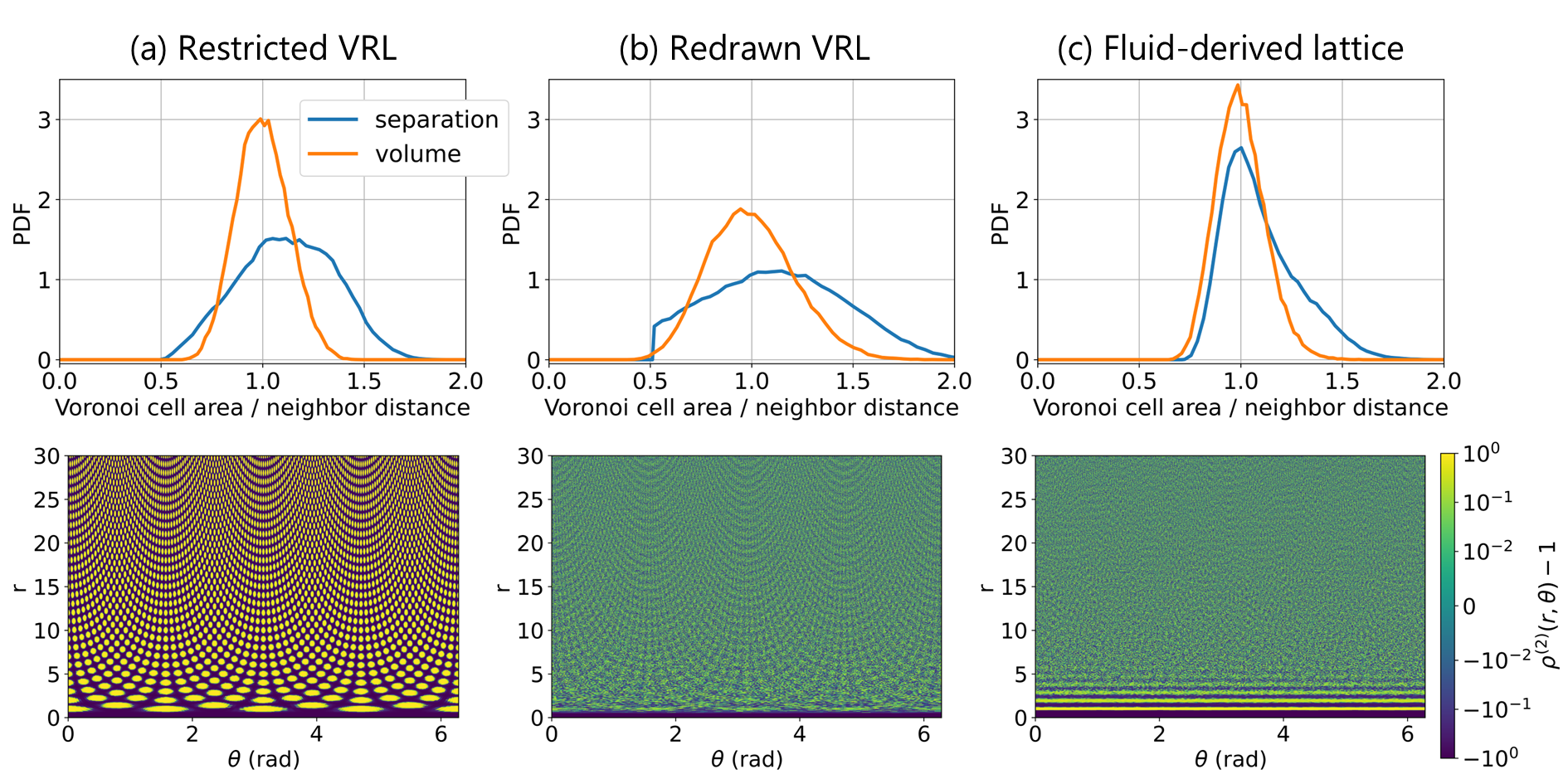}
    \caption{\textbf{Vectorizable random lattices have orientational order.} The top row shows the probability density functions (PDFs) of the distances between neighbors (blue) and the volumes of the Voronoi cells associated to the lattice sites (orange). The bottom row gives the pair distribution functions $\rho^{(2)}$ of the lattice sites in polar coordinates on a symmetrical log-scale. For the two vectorizable random lattices (VRLs), we chose $l_0 = 0.5$.}
    \label{fig:lattice_stats}
\end{adjustwidth}
\end{figure}

Figure~\ref{fig:lattice_stats} shows PDFs and pair distribution functions calculated for representative instances of the three disordered lattice types. In these examples, $l_0 = 0.5$ was used for both versions of the VRL. We note that the redrawn VRL gives broader distributions for both the separation between neighbor sites and the volume of the Voronoi cells compared to the restricted VRL and the fluid-derived lattice (top row). The redrawn VRL also has a sharp cutoff in neighbor separations at $l_0$. The pair distribution function (bottom row) of the restricted VRL shows clear anisotropy and long-range order. Although less pronounced, anisotropy and long-range order are also present in the redrawn VRL. The pair distribution function of the fluid-derived lattice shows no clear $\theta$ dependence, indicating that the lattice is isotropic. In addition, the fluid-derived lattice shows several oscillations at low $r$, indicating some short-range structure. These oscillations are typical for dense fluids and correspond to coordination layers around particles. These layers are related to the geometry of densely packed disks and would therefore also be expected in a 2D colony of disk-like bacteria. We conclude that only the fluid-derived lattice satisfies all requirements.

Next, we investigated whether the lattices can help resolve the artifacts we observed above. In Fig~\ref{fig:grow_square_diag}a we saw the diamond-shaped artifact in the nutrient constrained case. This suggests that lattice artifacts present in the solution to the reaction-diffusion equation governing the nutrient field can manifest in the overall colony shape. Therefore, we first looked at the continuous part of the model separately. We solved the bacterial growth model proposed by Kitsunezaki~\cite{kitsunezaki_interface_1997}, as this is a well-established model that can serve as a test case. The Kitsunezaki model consists of three density fields: the density of active motile bacteria $b$, of inactive bacteria $s$, and of a nutrient $n$. These fields are governed by three reaction-diffusion equations given by
\begin{align}
    \frac{\partial b}{\partial t} &= \nabla (D_0 b \nabla b) + nb - \mu b, \label{eq:kits_b}\\
    \frac{\partial n}{\partial t} &= \nabla^2 n - bn, \\
    \frac{\partial s}{\partial t} &= \mu b,
\end{align}
where $\mu$ is the rate of bacterial differentiation to the inactive state and $D_0$ the bacterial diffusion coefficient. The non-linear diffusion term in Eq~\eqref{eq:kits_b} is intended to model variable bacterial motility due to production of lubricant by the bacteria themselves. We set $D_0 = 0.1$, $\mu = 0.15$, and the initial nutrient concentration to $n_0 = 1$ everywhere. Our initial condition for the $b$ field was a disk of radius 5 with $b = 1$ placed in the center of the simulation domain. Everywhere else we set $b=0$. Finally, we initialized $s = 0$ everywhere. We used zero-flux boundary conditions for all fields.

We solved the Kitsunezaki model 500 times on each of the lattices to quantify any potential lattice artifacts with the convex-hull method described in the Methods section. Each simulation was performed on a new realization of the VRLs with $l_0 = 0.5$. We also attempted to solve the Kitsunezaki model for some $l_0 < 0.5$. However, this resulted in numerical instability for $l_0 \leq 0.4$, originating from the spatial derivatives in the diffusion terms. We randomly shifted the fluid-derived lattice such that the initial disk of the $b$ field was placed at a different position on the fluid-derived lattice each time.

\begin{figure}[!ht]
\begin{adjustwidth}{-2.25in}{0in}
    \centering
    \includegraphics[scale = 1.15]{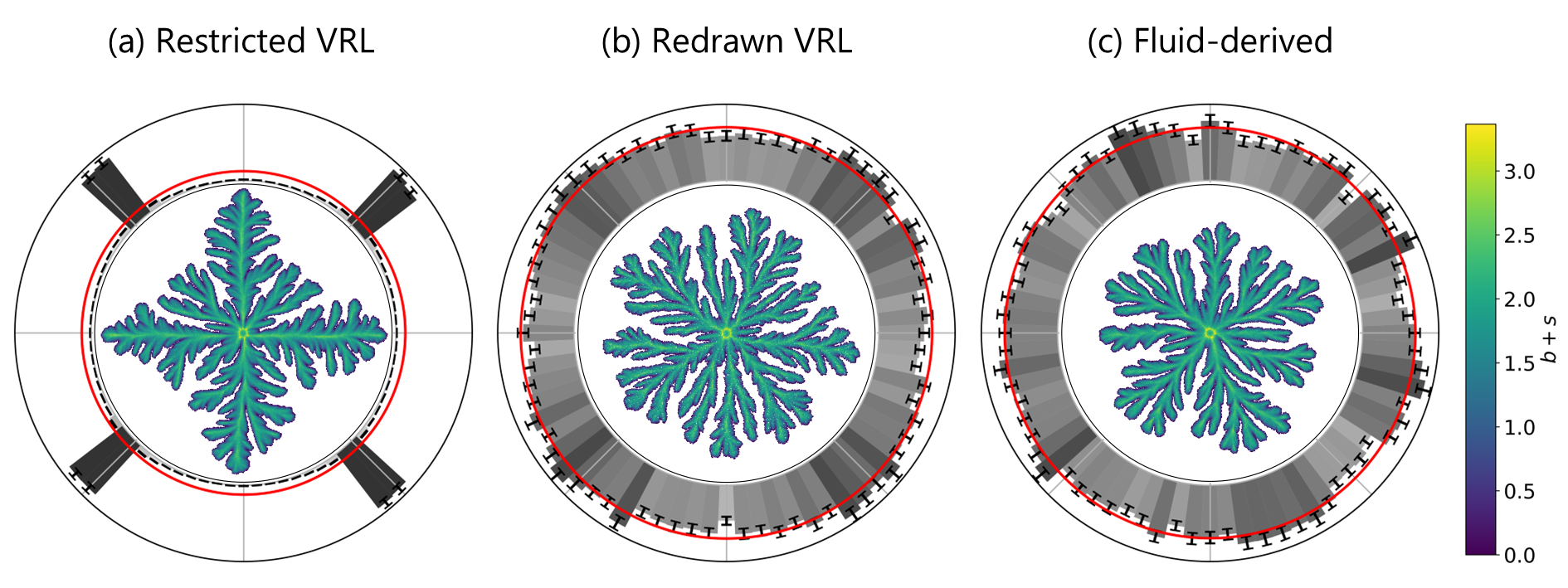}
    \caption{\textbf{Solutions to a continuous bacterial growth model on the vectorizable random lattices show discrete orientational symmetry, while solutions on the fluid-derived lattice do not.} The Kitsunezaki model~\cite{kitsunezaki_interface_1997} was solved on the three disordered lattices with parameters $D_0 = 0.1$ and $\mu = 0.15$. The initial condition for the density of active motile bacteria $b$ was a disk of radius 5 with $b=1$ and $b=0$ everywhere else. The nutrient density field $n$ was initialized with $n=1$ everywhere. The minimum distance between lattice sites in the VRLs $l_0$ was set to 0.5. The central figures show a single representative result of $b + s$ where $s$ is the density of inactive bacteria. The outer rings show the distribution of normal vectors to the fitted convex hulls as in Fig~\ref{fig:grow_square} based on 500 (independent) realizations.}
    \label{fig:cont_convhull}
\end{adjustwidth}
\end{figure}

A representative final colony and the normal vector distributions are plotted in Fig~\ref{fig:cont_convhull}. We see a pronounced diamond shape in the solutions on the restricted VRL in Fig~\ref{fig:cont_convhull}a. This diamond shape is much less pronounced in the colonies resulting on the redrawn VRL (Fig~\ref{fig:cont_convhull}b) and hard to recognize in individual instances. However, the histogram does reveal the same four-fold symmetry that we recognize from our results for the agent-based colonies with the diamond-shaped artifact. These results show that some of the structure of the square reference lattice used to generate the VRLs is present in the solutions to the Kitsunezaki model. On the other hand, the solutions on the fluid-derived lattice in Fig~\ref{fig:cont_convhull}c show no discrete orientational symmetry.

\begin{figure}[!ht]
\begin{adjustwidth}{-2.25in}{0in}
    \centering
    \includegraphics[scale = 1.15]{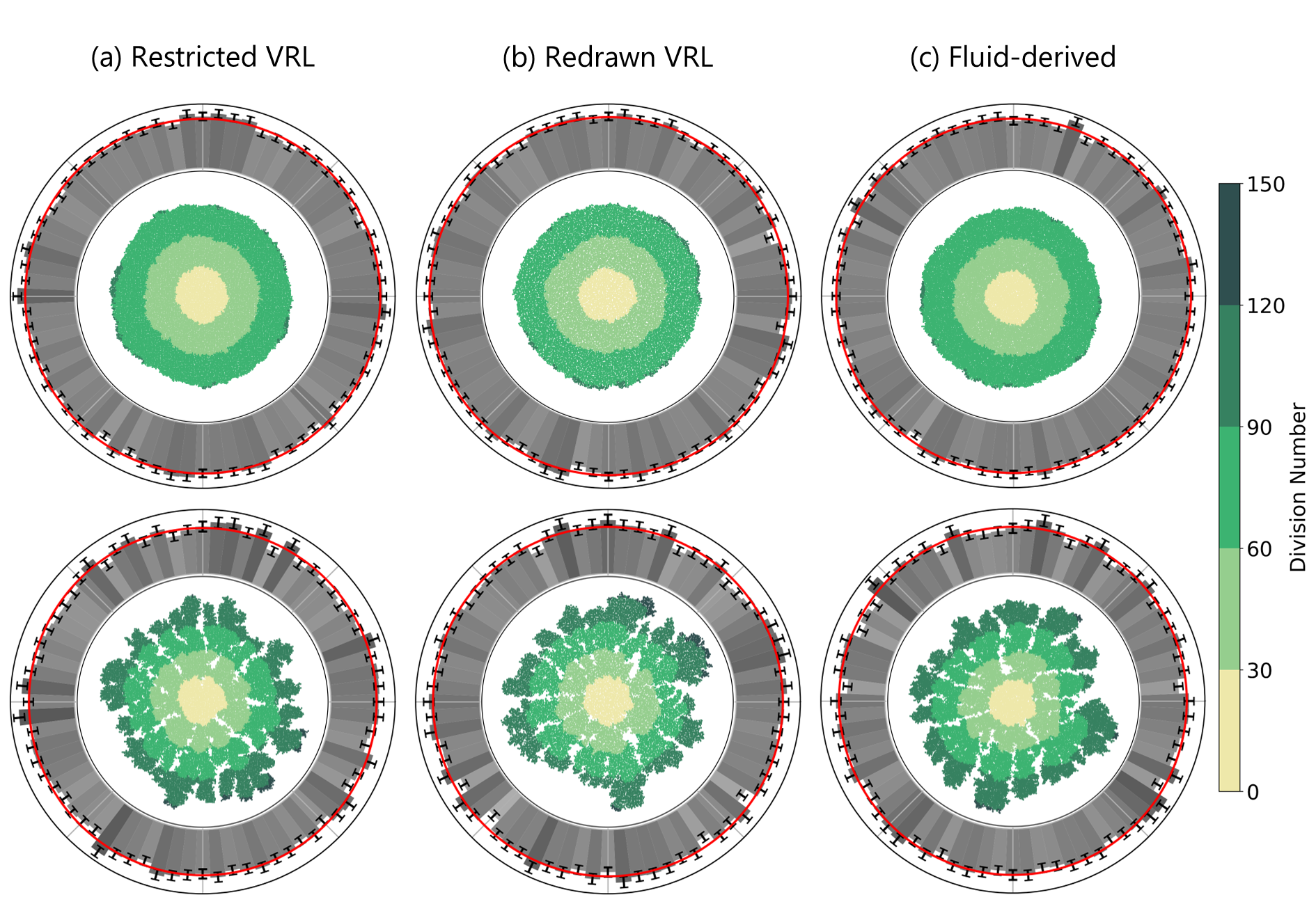}
    \caption{\textbf{The hybrid model with various disordered lattices does not display lattice artifacts.} The top row shows simulations with high initial nutrient concentration ($c_0 = 3$) and the bottom row shows simulations with low initial nutrient concentrations ($c_0 = 0.7$). The minimum separation between lattice sites in the VRLs $l_0$ was set to 0.5. The selection of parameters and representation is otherwise the same as in Fig~\ref{fig:grow_square}.}
    \label{fig:hybrid_convhull}
\end{adjustwidth}
\end{figure}

Next, we investigated how well the minimal hybrid model works on the various disordered lattices. We performed simulations with both high and low $c_0$ and applied the convex-hull procedure for each $c_0$ on all disordered lattices. Representative final colony shapes and the histograms of normal vectors to the fitted convex hulls are given in Fig~\ref{fig:hybrid_convhull}. We see that all three disordered lattices are effective in removing the discrete orientational symmetry observed in Figs~\ref{fig:grow_square},~\ref{fig:grow_square_diag}, and~\ref{fig:cont_convhull}.

A key distinction between the continuous and agent components of the hybrid model lies in how they interact with the underlying lattice. The agent component of the model has a binary neighbor relation: a site is a neighbor or it is not. The continuous component  of the model accounts for distances between sites, sizes of Voronoi cell walls, and Voronoi cell areas. This difference between the continuous and agent components of the hybrid model suggests that an agent-based model that includes more aspects of the lattice than the agent component of our hybrid model would be affected by lattice artifacts caused by the VRLs.

To test this hypothesis, we formulated a model of motile bacteria with cell-to-cell adhesion. We start with a fixed number of bacteria arranged in a colony. The energy $H$ of a configuration of bacteria is determined by the adhesion between bacteria and is given by:
\begin{align}
    H = -\frac{1}{2}  \sum_{i=1}^{N_b}\sum_{j \neq i}^{N_b}J L_{ij},
    \label{eq:adhesion_energy}
\end{align}
where $N_b$ is the number of bacteria, $J$ determines the strength of adhesion per unit contact length between bacteria, and $L_{ij}$ is the length of the Voronoi cell wall between the sites occupied by bacteria $i$ and $j$. Note that $L_{ij} = 0$ when bacteria $i$ and $j$ are not on neighboring lattice sites. Equation~\eqref{eq:adhesion_energy} says that the energetically most favorable position for a bacterium is to be surrounded by other bacteria on all sides. We then use a Monte Carlo method described in the Methods section to find the equilibrium configuration of the bacterial colony.

\begin{figure}[!ht]
\begin{adjustwidth}{-2.25in}{0in}
    \centering
    \includegraphics[width=1.0\linewidth]{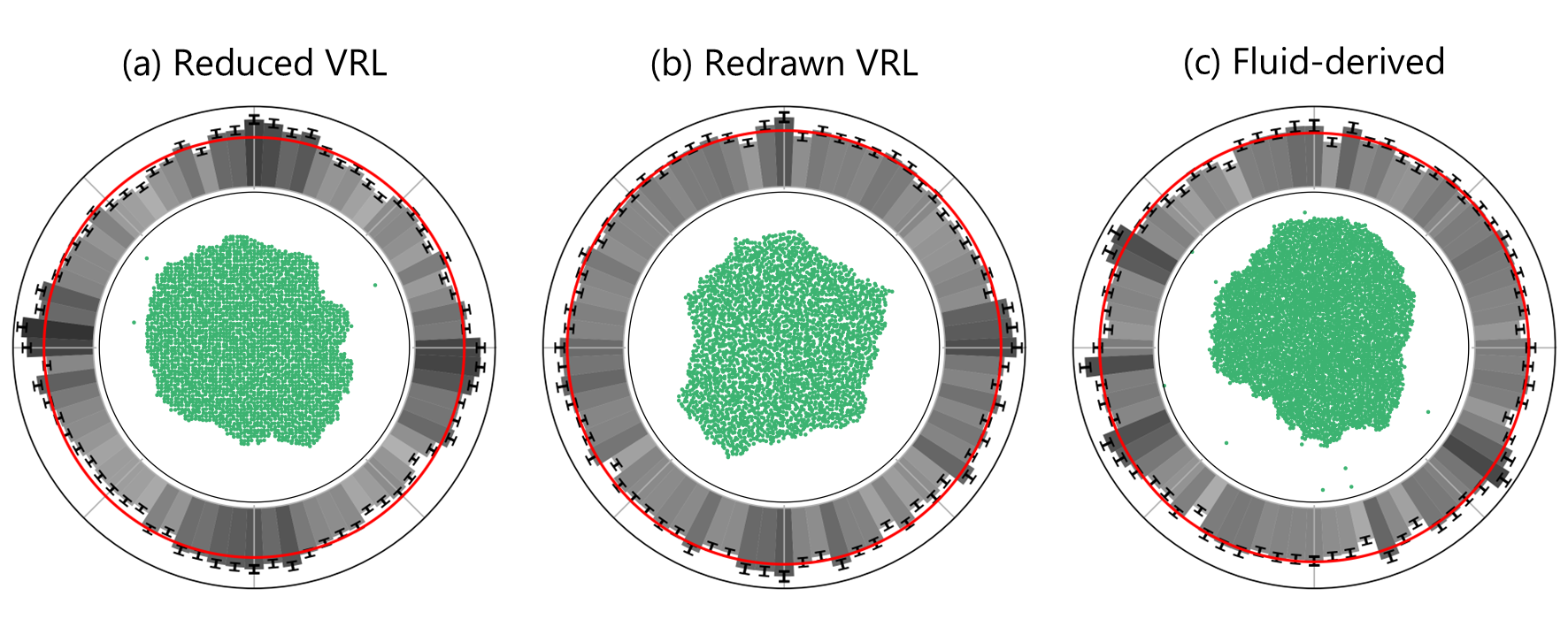}
    \caption{\textbf{Including adhesion between cells can cause lattice artifacts to reappear in some realizations of the disordered lattice.} The adhesion strength per unit contact area between cells was set to $J=4$. The central figures (green) show a representative final colony shape. The representation of the outer rings constructed from 1000 repetitions is as in Fig~\ref{fig:grow_square}.}
    \label{fig:adhesion_convhull}
\end{adjustwidth}
\end{figure}

We performed simulations of this model using the three disordered lattices. We characterized the resulting equilibrium configurations by fitting convex hulls to the main cluster as detailed in the Methods section. We performed simulations starting with a circular colony of radius $r = 30$ and starting with a square colony with edge length $\sqrt{\pi}r$ for each of the lattices and ran the simulations long enough such that the convex hull normal vector histograms resulting from the square and the circular initial conditions became indistinguishable. The combined results of the 1000 simulations on each lattice are given in Fig~\ref{fig:adhesion_convhull}. The lattice-induced four-fold symmetry has reappeared on the restricted VRL in Fig~\ref{fig:adhesion_convhull}a as expected. We do not detect this lattice artifact on either the redrawn VRL or the fluid-derived lattice of Figs~\ref{fig:adhesion_convhull}b and ~\ref{fig:adhesion_convhull}c, respectively. This shows that the geometric properties of the lattice are indeed important for the formation of the discrete orientational symmetry.

\subsection*{\label{sec:model_performance}Numerical performance of our implementation}

\begin{figure}[!ht]
    \centering
    \includegraphics[width=0.9\linewidth]{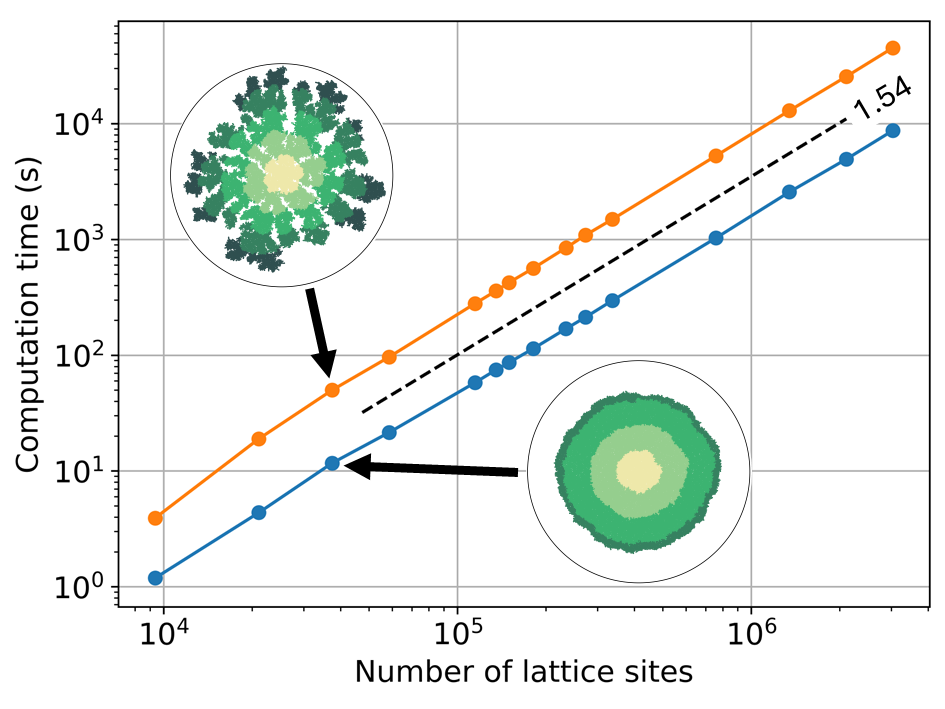}
    \caption{\textbf{Computation time shows power-law scaling of our hybrid bacterial growth model with the number of lattice sites.} The scaling was determined both for simulations with high initial nutrient concentration (blue dots) and for simulations with low initial nutrient concentration (orange dots). The connecting lines are to guide the eye. The exponent of the power law fit is $ 1.54 \pm 0.01$ (dashed line). The longest running simulation consisted of $3,034,566$ lattice sites.}
    \label{fig:performance_scaling}
\end{figure}

We have seen that our fluid-based method can overcome lattice-artifact issues. Let us now turn to the performance of our implementation. We performed a series of simulations with different system sizes to understand the scaling of our method. We ran the simulation three times for each system size and recorded the mean system time when the mean nutrient concentration was lower than 0.01 or at least 70\% of the lattice sites were occupied. All simulations were run on a workstation equipped with an Intel Core i7-7700 CPU running at a maximum frequency of 4.2 GHz, and 16 GB of system memory. The results are plotted in Fig~\ref{fig:performance_scaling}.

The hybrid model exhibits power-law scaling with system size, with a scaling exponent of $1.54 \pm 0.01$. There is some deviation from power-law scaling at smaller system sizes ($< 3 \cdot 10^4$ lattice sites), primarily due to the Voronoi tessellation performed at the start of each simulation. The largest system tested, with low initial nutrient concentration, contained $3,034,566$ lattice sites with a final colony consisting of $1,639,299$ bacteria. This simulation took approximately 12.5~hours to compute and corresponds to a real-world colony about 2~mm in diameter (assuming bacteria that are 1 $\mu$m in diameter). Computation time can be further reduced by parallelization, especially since the continuous component of the model is well-suited for this.

In order to see how the performance of the hybrid lattice-based model with fluid-derived lattice compares to that of a more detailed off-lattice agent-based method, we performed benchmark simulations using the \textbf{\texttt{iDynoMiCS}} software package~\cite{lardon_idynomics_2011}. This software package was developed to simulate bacterial biofilms in a flow cell. For the \textbf{\texttt{iDynoMiCS}} simulations, we used the simulation parameters as in Ref~\cite{young_lineage_2022}. In the off-lattice agent-based simulations, we set the bulk nutrient concentration of the limiting nutrient $S_{\mathrm{bulk}}$ to $10^{-2}$~g/L to model dense, non-branching colonies. We initialized the simulations by seeding 200 bacteria at the bottom of the rectangular simulation domain. The lateral boundaries of the simulation domain are periodic and the lower boundary is a wall. The upper boundary is connected to the bulk nutrient reservoir. Half the initial bacteria were colored blue and the other half red. Bacteria passed their color on to their daughter cells.

\begin{figure}[!ht]
    \centering
    \includegraphics[width=0.9\linewidth]{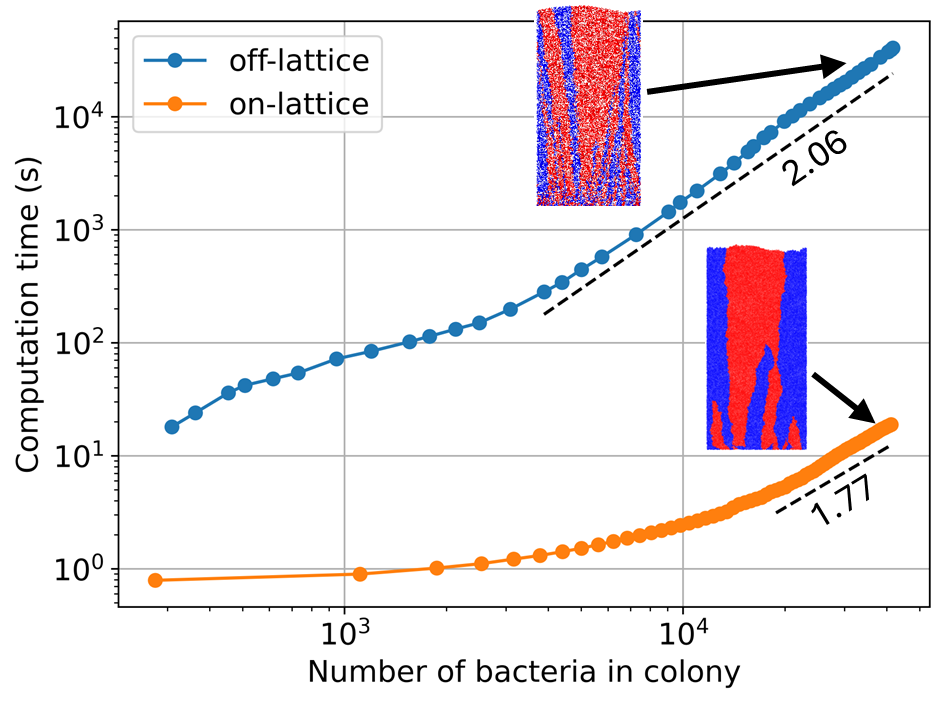}
    \caption{\textbf{The hybrid on-lattice model can be computed considerably faster than off-lattice methods.} Computation time of the hybrid lattice-based model (orange dots) is compared with an agent-based off-lattice simulation of a similar system (blue dots). The connecting lines are to guide the eye. Colony size is plotted against computation time during a single simulation. A power law was fitted to both curves (dashed lines). The exponent fitted to the off-lattice curve was $2.0681 \pm 0.0002$ and the exponent fitted to the on-lattice curve was $1.7651 \pm 0.0001$. The off-lattice simulation was performed using the \textbf{\texttt{iDynoMiCS}} software package~\cite{lardon_idynomics_2011} with parameters chosen as in Ref~\cite{young_lineage_2022} and with bulk limiting nutrient concentration $S_{\textrm{bulk}} = 10^{-2}$ g/L. The parameters used in the hybrid lattice-based simulation are initial nutrient concentration $c_0 = 3.0$ and the remaining parameters as in Table~\ref{tab:parameters}. The insets show the final colonies where half the initial bacteria were colored red and the other half blue. Color was inherited from parent to daughter cells. The final configuration of the off-lattice simulation contains 41,697 agents and took approximately 11 hours to compute. The final configuration of our hybrid on-lattice simulation contains 42,015 agents and took approximately 19 seconds to compute.}
    \label{fig:idyno}
\end{figure}

We adapted our hybrid lattice-based model to feature a rectangular simulation domain to better match the environment simulated by \textbf{\texttt{iDynoMiCS}}. The left and right boundaries are periodic and the top and bottom are walls with zero-flux boundary conditions as before. We initialized the hybrid model by seeding 200 bacteria at the bottom of the simulation domain and we set the initial nutrient concentration $c_0$ to 3.0 to model dense, non-branching colony morphologies. The red and blue colors are the same as in the \textbf{\texttt{iDynoMiCS}} simulations. The remaining parameters are as in Table~\ref{tab:parameters}. We tracked the computation time as the colonies increased in size. These results of the agent-based and lattice-based simulations can be found in Fig~\ref{fig:idyno}.

Our hybrid lattice-based mode for bacterial growth reaches similar colony sizes in significantly (often orders-of-magnitude) less time than the off-lattice agent-based growth model implemented by \textbf{\texttt{iDynoMiCS}}. Our hybrid model grew to $42,015$ bacteria in approximately 19 seconds, while the \textbf{\texttt{iDynoMiCS}} simulation reached $41,697$ bacteria in about 11 hours. We also compared simulations in the branching regime. However, the two simulation methods resulted in different colony morphologies. We discuss this discrepancy in further detail in~\nameref{S1_Appendix}.

There are some caveats to the above comparison of computation time. For example, \textbf{\texttt{iDynoMiCS}} is written in \textbf{\texttt{Java}}, whereas our hybrid lattice-based model is implemented in \textbf{\texttt{C++}}. Furthermore, some of the physics included in \textbf{\texttt{iDynoMiCS}}, like the shoving interaction between bacteria, is absent in the minimal hybrid model. However, the significant difference in computation time suggests it is worthwhile to consider whether a problem of interest can be studied with lattice-based methods instead of a more detailed agent-based model.

\section*{Discussion and conclusion}

To characterize lattice artifacts in simulations of growing bacterial colonies we have focused on a broad class of hybrid lattice-based models. Previous studies utilizing similar hybrid models used either a square lattice~\cite{gerlee_hybrid_2008, van_dijk_silico_2016} or a hexagonal lattice~\cite{borer_spatial_2020}. In our investigation of the minimal hybrid model using a square lattice, we observed a distinct four-fold orientational symmetry in the shape of resulting colonies, resembling patterns seen in the Eden model~\cite{eden_two-dimensional_1961, batchelor_limits_1991}. This prompted us to explore whether disordered lattices could eliminate such artifacts.

We found that all three disordered lattices that we considered, successfully resolved the lattice-induced orientational symmetry in the hybrid agent-based model. We had anticipated that some degree of orientational symmetry would remain in the restricted vectorizable random lattice (VRL), based on the following two observations. (i)~We saw significant orientational symmetry in solutions to a continuous model on this lattice. (ii)~The restricted VRL itself maintains four-fold orientational symmetry. A key distinction between the continuous and agent components of the minimal hybrid model lies in their use of the underlying lattice: the continuous component includes the distance between sites, the area of Voronoi cells, and the length of Voronoi cell walls, while the agent component only requires a binary neighborhood relationship. It is clear that the disordered lattices can effectively establish the appropriate binary neighborhood relationships for the agent component, where the weighted Moore neighborhood on the square lattice failed to do so. Further investigation is needed to determine the exact properties that these neighborhood relations should exhibit to suppress artifacts.

The four-fold orientational symmetry reappeared when we introduced a geometric property of the lattice in the agent component of the model of motile bacteria with cell-to-cell adhesion. From this, we conclude that the impact of symmetries in the underlying lattice strongly depends on the specific model used. Therefore, one should not assume that a specific choice of lattice will suffice, and it is essential to explicitly check for relevant lattice artifacts.

It is important to note that any lattice will impose some structure on numerical modeling results, which may not be suitable for the system of interest. The fluid-derived lattice can be adapted to better suit the characteristics of the model system. For instance, constructing the lattice using simulations of a fluid of ellipsoids might be more appropriate to modeling a system of rod-shaped bacteria. Another possible approach would be to base the lattice on an experimental realization of the system of interest. In this case, the lattice would be reconstructed from measurements of the structure of the experimental result.

Let us now turn to the efficiency of our approach. One motivation for our study was to simulate large systems, a task traditionally achievable through continuous models. For instance, the continuous model of Schwarcz~\textit{et al.}~\cite{schwarcz_uniform_2016} successfully simulated colonies up to 5~cm in radius (equivalent to approximately $10^{10}$ cells~\cite{nelson_penicillin_2000}), demonstrating the feasibility of modeling large-scale systems. However, the increased availability of observations of individual microbes has significantly advanced our understanding of bacterial behavior at the individual level~\cite{stewart_physiological_2008, hellweger_bunch_2009}. Incorporating these individual-level properties into continuous models presents a challenge. Furthermore, when simulations are applied to study evolutionary processes, the discrete nature of individual bacteria becomes crucial.

Agent-based methods are more appropriate than continuous models to tackle questions involving inheritance and diversity, as they keep track of individuals. However, simulating the large system sizes needed for such studies with such methods presents significant challenges. This is especially the case when using off-lattice approaches, which have to also propagate the equations of motion. Typically, off-lattice agent-based systems consist of 10,000 to 100,000 agents~\cite{rana_spreading_2017, los_time_2024, farrell_mechanically_2013}. Some studies have managed to simulate larger systems by optimizing the model and not actively simulating every individual agent. For example, Young~\textit{et al.}~\cite{young_active_2023} were interested in the effect of active layer dynamics on the morphology of growing biofilms. They accelerated their simulations by removing inactive cells far behind the growing front. This allowed them to reach system sizes of $10^5$ bacteria. The \textbf{\texttt{NUFEB}} code, developed by Li~\textit{et al.}, is notable for enabling the simulation of exceptionally large bacterial populations, with up to $10^7$ individuals~\cite{li_nufeb_2019}. This implementation builds on the parallelization offered by the popular simulation packages \textbf{\texttt{LAMMPS}} and \textbf{\texttt{OpenFOAM}}, but can only simulate such large systems effectively by making use of considerable computational resources.

A method by which large numbers of bacteria can be simulated with genetic inheritance, but without the need for high-performance computing, is lattice-based modeling. Such approaches have been considered for system sizes in excess of 100,000 bacteria~\cite{mobius_how_2015}. Mukhamadiarov~\textit{et al.}~\cite{mukhamadiarov_clonal_2024} even considered a colony consisting of $1.6\times 10^7$ bacteria. Our approach stands out from these, as we use disordered lattices as opposed to regular lattices, which removes discrete orientational symmetries. We also considered the efficiency of the lattice-based approach compared to a full off-lattice simulation. Our results showed that the fluid-lattice-based hybrid model is orders of magnitude faster than a similar off-lattice simulation implemented by \textbf{\texttt{iDynoMiCS}}. This makes it clear that simulations involving millions of bacteria are accessible in hours on a regular desktop machine.

In summary, we have studied a minimal hybrid model for bacterial populations that constrains individual bacteria a lattice and couples them to a continuous field. We were able to simulate millions of individual bacteria in a matter of hours on a regular desktop computer. This is far beyond what off-lattice codes can do, as we have argued above. In addition, we demonstrated that fluid-derived disordered lattices can be effective in mitigating lattice-induced symmetries. In future work, the hybrid model can be further extended to include features like active motility, lubricant production, and anisotropic bacteria. These extensions could help shed light on the formation of chiral colonies~\cite{ben-jacob_holotransformations_1994, cohen_orientation_2000} and colonies consisting of concentric rings~\cite{rafols_formation_1998, wakita_self-affinity_1997}. The larger accessible population sizes and the individual nature of the bacteria in our model will allow us to apply our model to questions relating to evolutionary processes in spatially structured populations. One example of this is selection in spatially structured populations due to a spatially inhomogeneous selection pressure, which could be relevant in the evolution of antimicrobial resistance~\cite{greulich_mutational_2012, baymSpatiotemporalMicrobialEvolution2016, hermsen_adaptation_2016}. We hope that the hybrid approach on a disordered lattice will prove to be a useful tool for the community interested in modeling spatially structured populations.

\section*{\label{sec:methods}Methods}

\subsection*{\label{sec:method_hybridmodel}Definition of the hybrid bacterial-growth model}

Here, we will provide details of the hybrid bacterial-growth model used in Figs~\ref{fig:grow_square},~\ref{fig:diag_weighted},~\ref{fig:grow_square_diag},~and~\ref{fig:hybrid_convhull}. The bacteria are modeled as individual agents at positions $\vec{x}_i$. The nutrients are modeled as a continuous field $c$ governed by a reaction-diffusion equation:
\begin{align}
    \frac{\partial c}{\partial t} = D \nabla^2 c - \sum_{i=1}^{N_\textrm{b}}\delta(\vec{x}_i - \vec{x}) f(c).
    \label{eq:nutrientfield}
\end{align}
The first term describes the diffusion of nutrient, with diffusion constant $D$. The second term describes the consumption of nutrient by all $N\ped{b}$ bacteria. The bacteria appear as point sinks, implemented by Dirac delta peaks $\delta(\vec{x})$ at the center of each bacterium. The function $f(c)$ describes the concentration-dependent nutrient uptake rate, which we chose to be specified by the Monod equation:
\begin{align}
    f(c) = v_\mathrm{max} \frac{c}{K + c},
    \label{eq:monod}
\end{align}
with $v_\mathrm{max}$ the maximum nutrient consumption rate and $K$ the half-saturation constant. The Monod equation~\eqref{eq:monod} states that, at low nutrient concentrations, the nutrient consumption rate increases with the nutrient concentration. When the nutrient concentration is high, bacteria consume the nutrients at a fixed maximum rate $v_\textrm{max}$.

The positions of the bacteria  $\vec{x}_i$ are constrained to a two-dimensional (2D) lattice. Each lattice site can be occupied by at most one bacterium. Each bacterium is assigned an internal state $n$ which describes the amount of nutrients consumed. Therefore, the rate of change of $n$ is also given by the Monod equation~\eqref{eq:monod}. When $n$ exceeds a threshold $n_\mathrm{g}$, the bacterium has consumed enough nutrient to produce one daughter cell. This daughter cell is randomly placed on an empty neighbor lattice site. If there are no unoccupied neighbor sites, the bacterium is unable to produce the daughter cell. The daughter cell receives $n_d$ nutrients where $n_d$ is chosen from the interval $[\frac{n}{2}-\delta, \frac{n}{2}+\delta)$ with uniform probability. We set $\delta = 0.2$ for all simulations. The parent cell retains $n-n_d$ nutrients. At every time step, the order in which these division events are executed is randomized.

\subsection*{\label{sec:methods_latticegen}Generating a disordered lattice}

In the Results section, we evaluated three types of disordered lattice: reduced vectorizable random lattices (VRLs) as proposed by Moukarzel and Herrmann~\cite{moukarzel_vectorizable_1992}, redrawn VRLs as proposed by Tucker~\cite{tucker_new_2010}, and fluid-derived random lattices. We generated the VRLs as described in the Results section (Fig~\ref{fig:lattice_gen}). The fluid-derived random lattices were obtained by molecular-dynamics simulations of a dense (2D) fluid. We simulated 40,000 soft disks (bidisperse sample with stoichiometry 1:1 and size ratio 1.25) in a square 2D domain with periodic boundary conditions at area fraction $\varphi = 0.7$. The interaction between the disks is governed by the WCA potential~\cite{weeks_role_1971}:
\begin{align}  
    u_{\textrm{WCA}}(r_{ij}) = \left\{
    \begin{IEEEeqnarraybox}[][c]{l?r}
    \IEEEstrut
    4 \epsilon \left[ \left(\frac{\sigma_{ij}}{r_{ij}} \right)^{12} - \left(\frac{\sigma_{ij}}{r_{ij}} \right)^6 \right] + \epsilon,  \quad & \text{if } r_{ij} < 2^{1/6}\sigma_{ij},  \\
    0, & \text{if } r_{ij}\geq 2^{1/6}\sigma_{ij},
    \IEEEstrut
    \end{IEEEeqnarraybox}
    \right.
    \label{eq:WCA}
\end{align}
where $\epsilon$ is the interaction strength, $r_{ij}$ the distance between particles $i$ and $j$, and $\sigma_{ij} = (d_i + d_j) / 2$ with $d_i$ the diameter of particle $i$. We ran the simulations with $k_b T/\epsilon = 2.5$ where $k_b$ is the Boltzmann constant and $T$ the temperature. We used the center of mass coordinates of the disks as pseudo-random coordinates for our lattice. When we needed more than 40,000 lattice sites, we extended the lattice by copying the original 40,000 coordinates until our lattice had the desired size.

From our set of pseudo-random coordinates, we obtained the desired disordered lattice by connecting neighboring sites. Here, neighbors of a lattice site are defined using a Delaunay triangulation. This is constructed from the regular Voronoi tessellation corresponding to our pseudo-random coordinates. The Voronoi tessellation partitions the domain into cells, where cell $i$ is the set of all points that are closer to lattice site $i$ than to any other site. Here, we use the \textbf{\texttt{Voro++}} library developed by Rycroft\cite{rycroft_voro_2009} to construct the tessellation. By connecting all the sites that share a Voronoi cell wall, we obtain the associated Delaunay triangulation, which will serve as our disordered lattice. Two sites are neighbors if they are directly connected in the Delaunay triangulation. We rescaled the coordinates in the final configurations such that the average distance between sites equals the unit of distance $\Delta x$.

\subsection*{\label{sec:method_integration}Numerical integration of reaction-diffusion equations}

The reaction-diffusion equation~\eqref{eq:nutrientfield} is discretized using the same lattice that constrains the bacteria. We always use the forward Euler method for time integration and use circular zero-flux boundary conditions. We use different approaches to calculate the spatial derivatives of Eq~\eqref{eq:nutrientfield}, that depend on the specifics of the lattice.

For square lattices, we calculate the gradient of the nutrient field using a finite differences scheme. We only consider the von Neumann neighborhood of each lattice site, resulting in a five-point central differences scheme. The discretized version of Eq~\eqref{eq:nutrientfield} thus becomes
\begin{align}
   c_{i,j}^{k+1} = c_{i,j}^k + \frac{\Delta t}{A} \left[D \left(c_{i-1,j}^k + c_{i,j-1}^k - 4 c_{i,j}^k + c_{i+1,j}^k + c_{i,j+1}^k \right) - \mathbf{1}_\textrm{b}(\vec{x}_{i,j}) f(c_{i,j}^k)\right],
   \label{eq:fdm}
\end{align}
where $A = \Delta x^2$ is the area of a grid cell with lattice spacing $\Delta x$, $\Delta t$ the time-step size, $D$ the nutrient diffusion coefficient, $c_{i,j}^k$ the nutrient concentration at $\vec{x} = \vec{x}_{i,j} = (i \Delta x, j \Delta x)$ at time $t = k \Delta t$, and $\textbf{1}_\textrm{b}(\vec{x})$ the indicator function that equals 1 when a bacterium is present and 0 otherwise. The boundary conditions are implemented with the method proposed in Ref~\cite{hunt_finite_1978}.

For the disordered lattices, we use a finite-volumes approach~\cite{barth_finite_2017}. The lattice needs to be subdivided into control volumes each containing one lattice site. We use the Voronoi cells we calculated when constructing the lattice for this purpose. Then Eq~\eqref{eq:nutrientfield} discretized on the disordered lattice becomes
\begin{align}
    c_i^{k+1} = c_i^k +\frac{\Delta t}{A_i} \left[ D \sum_{j\neq i}^N L_{ij}\frac{c_j^k - c_i^k}{d_{ij}} - \mathbf{1}_\textrm{b}(\vec{x}_i)f(c_i^k) \right],
    \label{eq:fvm}
\end{align}
with $c_i^k$ the nutrient concentration on the lattice site labeled with index $i$ at time $t = k\Delta t$, $A_i$ the area of the Voronoi cell associated with lattice site $i$, $N$ number of lattice sites, $D$ the nutrient diffusion coefficient, $L_{ij}$ the length of the Voronoi cell wall separating sites $i$ and $j$, which equals 0 when $i$ and $j$ are not neighboring sites, $d_{ij}$ the distance between sites $i$ and $j$, and $\Delta t$ the time-step size. The function $\mathbf{1}_\textrm{b}(\vec{x}_i)$ is the same as in Eq~\eqref{eq:fdm}.

\subsection*{\label{sec:methdods_mc}Monte Carlo simulations of motile bacteria with cell-to-cell adhesion}

The equilibrium configurations presented in Fig~\ref{fig:adhesion_convhull} were obtained as follows. The model assumes a fixed number of motile bacteria. The interaction between bacteria is given by cell-to-cell adhesion and is captured by the Hamiltonian of Eq~\eqref{eq:adhesion_energy}. We then use the Metropolis-Hastings Monte Carlo method~\cite{metropolis_equation_1953, hastings_monte_1970} to find equilibrium configurations. Trial moves consisted of attempts to move a randomly selected bacterium to a random neighbor site. If the neighbor site is occupied, the trial move is rejected. If not the trial move is accepted with probability
\begin{align}
    \textrm{acc}(o \rightarrow n) = \min\left[1, \exp\left({-\left(H(n)-H(o)\right) \frac{N_o}{N_n}}\right)\right],
\end{align}
where $o$ and $n$ indicate the old and new configuration respectively, $N_o$ and $N_n$ the number of neighbor sites of the selected bacterium in the old and new configurations, and $H$ the Hamiltonian governing the interaction between bacteria. The ratio of $N_o$ and $N_n$ is necessary to ensure detailed balance in disordered lattices where the number of neighbors is not the same for each site. Each Monte Carlo simulation consisted of $1.26 \times 10^{12}$ trial moves.

\subsection*{\label{sec:methods_convhull}Quantifying lattice-induced orientational symmetry}

We quantify the observed lattice artifacts by fitting a convex hull to each individual final colony shape. We use 500 samples of each simulation to gather statistics. Each simulation was performed on a different realization of the VRLs and initialized on a different position on the fluid-derived lattice. We use the \textbf{\texttt{SciPy}} package to fit the convex hulls~\cite{virtanen_scipy_2020}. Next, we determine the normal vectors to each segment of the fitted convex hull. These vectors are then used to construct a histogram of their angles, with each vector weighted by the length of its corresponding convex hull segment. The height of $k$-th bin in the histogram, denoted $h_k$, is then given by
\begin{align}
    h_k = \sum_{w\in W_k}w,
\end{align}
where $W_k$ is the set of all normal vector weights in bin $k$. The standard deviation $\sigma_k$ of the $k$-th bin is calculated as
\begin{align}
    \sigma_k = \sqrt{\sum_{w\in W_k} w^2}.
\end{align}
Finally, the histogram is normalized to yield a probability density. A schematic overview of this procedure is given in \nameref{S2_Figure}.

\section*{Acknowledgments}

We acknowledge Utrecht University for funding through the internal Science for Sustainability PhD programme. We are grateful to Hossein Nemati, Benjamin Planterose Jiménez, and Kim William Torre for useful discussions.

\begin{singlespace}
\section*{Author contributions}

\noindent\textbf{Conceptualization:} Joost de Graaf.\newline

\noindent\textbf{Formal analysis:} Bryan Verhoef.\newline

\noindent\textbf{Methodology:} Bryan Verhoef, Rutger Hermsen, Joost de Graaf.\newline

\noindent\textbf{Software:} Bryan Verhoef.\newline

\noindent\textbf{Supervision:} Rutger Hermsen, Joost de Graaf.\newline

\noindent\textbf{Validation:} Bryan Verhoef.\newline

\noindent\textbf{Visualization:} Bryan Verhoef.\newline

\noindent\textbf{Writing - original draft:} Bryan Verhoef.\newline

\noindent\textbf{Writing - review \& editing:} Rutger Hermsen, Joost de Graaf.

\end{singlespace}

\nolinenumbers

\begin{singlespace}
\bibliography{references}

\clearpage
\appendix
\section*{Supporting information}
\renewcommand{\thefigure}{S\arabic{figure}}
\setcounter{figure}{0}  

\input{S1_text.tex}

\clearpage
\input{S2_figure.tex}

\end{singlespace}

\end{document}

%% file: S1_text.tex
\renewcommand{\vec}[1]{\mathbf{#1}}

\paragraph{S1 Text} {\bf Differences in colony morphology in simulations with \textbf{\texttt{iDynoMiCS}} and the hybrid lattice-based method}
\label{S1_Appendix}
\newline

\begin{figure}[!ht]
    \centering
    \includegraphics[width=0.9\linewidth]{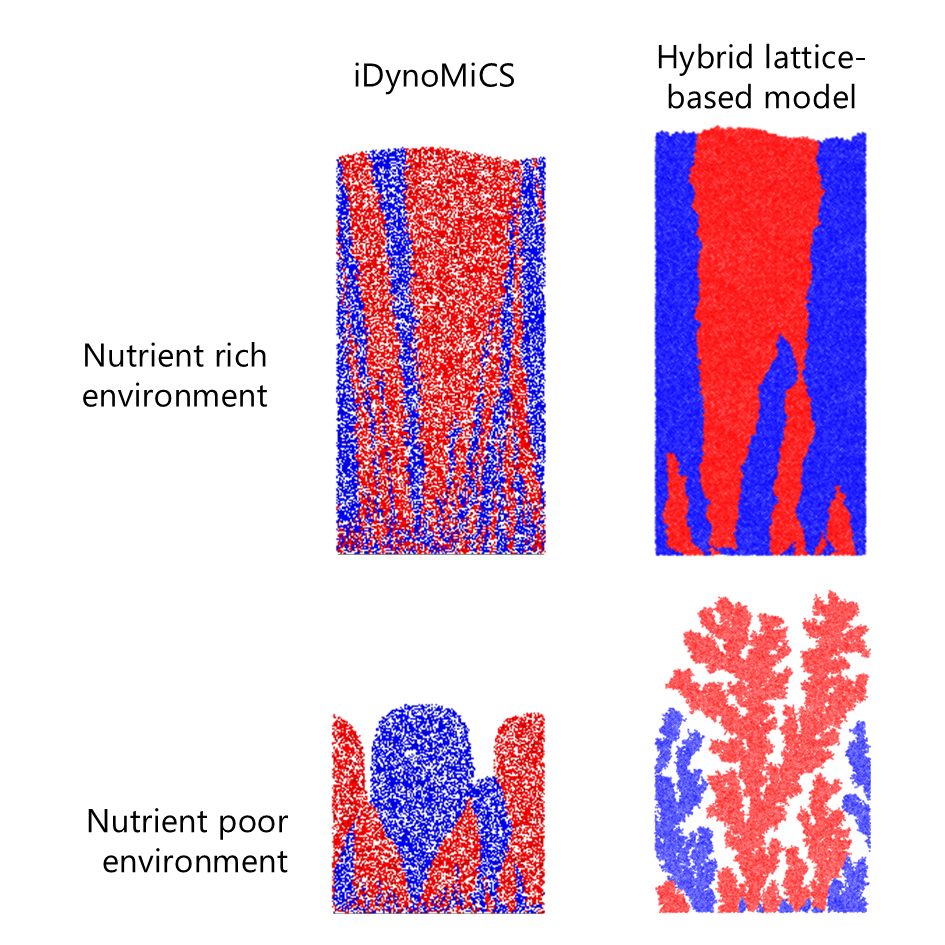}
    \caption{\textbf{The hybrid lattice-based model and \textbf{\texttt{iDynoMiCS}} result in different colony morphologies in nutrient poor environments.} Colonies grown under nutrient rich and nutrient poor environments using \textbf{\texttt{iDynoMiCS}}~\cite{lardon_idynomics_2011} and our hybrid lattice-based model are compared. Each simulation is seeded with 200 bacteria at the bottom of the simulation domain. Each bacterium is marked either red or blue with equal probability. Bacteria transmit this color to their daughter cells. For the \textbf{\texttt{iDynoMiCS}} simulations, the nutrient rich environment is set by the bulk limiting nutrient concentration $S_{\textrm{bulk}} = 10^{-2}$ g/L and the nutrient poor environment by $S_{\textrm{bulk}} = 10^{-3}$ g/L. The remaining parameter choices are as in Ref~\cite{young_lineage_2022}. For the hybrid lattice-based simulations, the nutrient rich environment is set by initial nutrient concentration $c_0 = 3$ and the nutrient poor environment by setting $c_0 = 0.7$. The remaining parameter choices are as in Table 1 in the main text.}
    \label{fig:idyno_shape}
\end{figure}

\noindent
As noted in the main text, we found that \textbf{\texttt{iDynoMiCS}}~\cite{lardon_idynomics_2011} and the hybrid lattice-based model give substantially different colony morphologies in nutrient constrained environments. Our hybrid lattice-based model gives a branched morphology where \textbf{\texttt{iDynoMiCS}} gives a fingered morphology in nutrient poor environments, as can be seen in the bottom row of Fig~\ref{fig:idyno_shape}. The \textbf{\texttt{iDynoMiCS}} model includes several features that our model does not. For example, \textbf{\texttt{iDynoMiCS}} models pressure-driven movement of bacteria and shoving between bacteria. Furthermore, the model system of \textbf{\texttt{iDynoMiCS}} is slightly different. Where we model a system akin to growth on an agar plate with a finite amount of nutrients, \textbf{\texttt{iDynoMiCS}} models a flow-cell system with a user-definable boundary layer and a nutrient concentration field fed by a bulk nutrient reservoir outside the boundary layer~\cite{lardon_idynomics_2011}. 

We compared \textbf{\texttt{iDynoMiCS}} to earlier work by Nadell~\textit{et al.}~\cite{nadell_emergence_2010} to get a better understanding of the most important differences between the various models. These authors model a similar flow-cell setup with a user defined boundary layer and a bulk solute reservoir and shoving interactions between agents. Despite these similarities to the system modeled by \textbf{\texttt{iDynoMiCS}}, Nadell~\textit{et al.} do find branching colonies at low nutrient availability as opposed to the fingering morphology found by \textbf{\texttt{iDynoMiCS}}~\cite{nadell_emergence_2010, bonachela_universality_2011}. The difference that stood out to us was the lack of pressure driven bacterial movement that is present in the work of Ref~\cite{nadell_emergence_2010, bonachela_universality_2011}. However, disabling this pressure-driven motion in \textbf{\texttt{iDynoMiCS}} results in a fingered morphology similar to the one that we observed earlier in Fig~\ref{fig:idyno_shape}. It is not clear to us which of the differences in the models causes the observed difference in colony morphology.

%% file: S2_figure.tex
\renewcommand{\vec}[1]{\mathbf{#1}}

\paragraph*{S2~Figure}
\label{S2_Figure}
{\bf Overview of the convex hull procedure for detecting orientational symmetries in the shape of simulated bacterial colonies}\newline

\begin{figure}[!ht]
    \begin{adjustwidth}{-2.25in}{0in}
    \centering
    \includegraphics[width=0.8\linewidth]{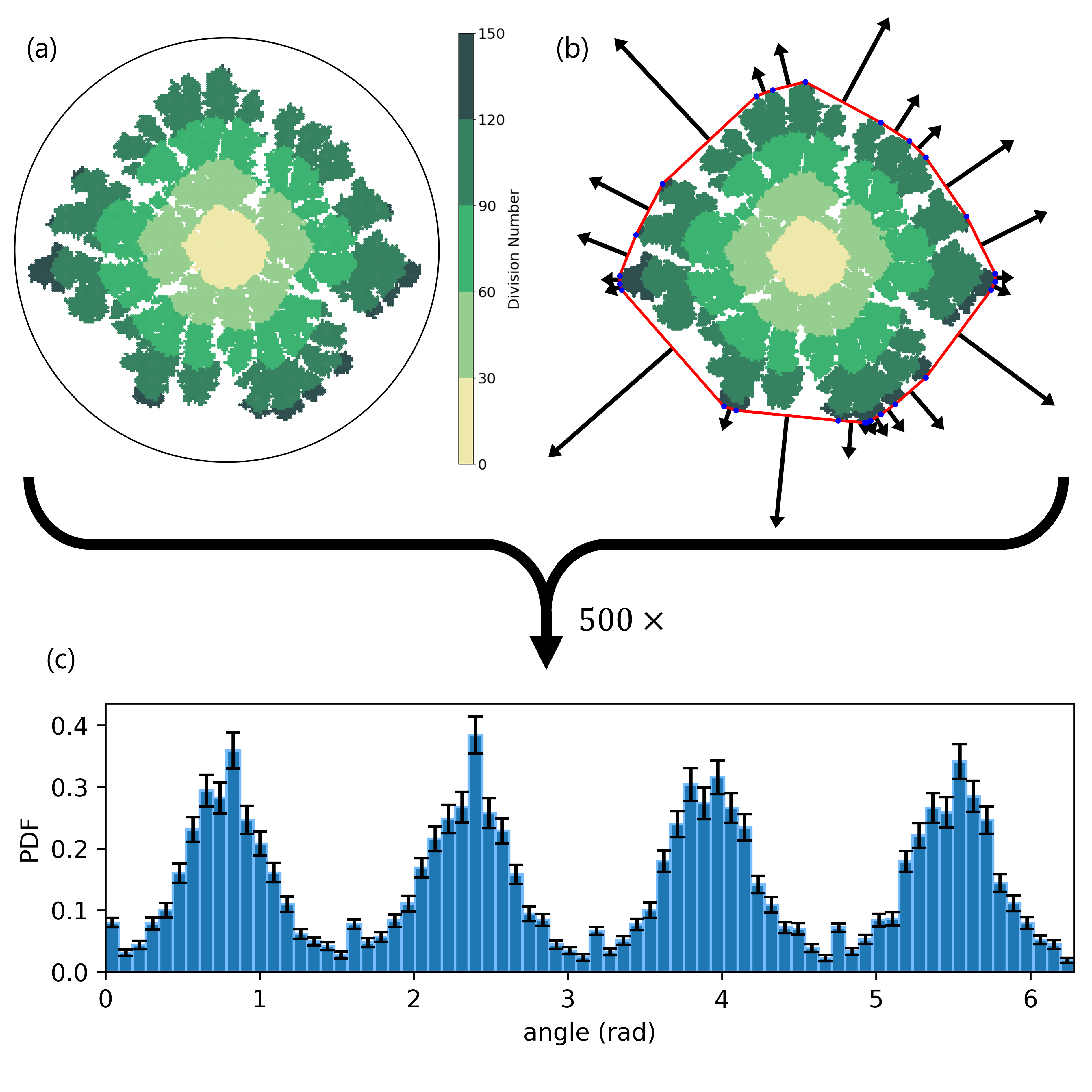}
    \caption{\textbf{Fitting of convex hulls to simulated colonies allows for quantification of lattice-induced symmetries.} (a) We perform simulations of the hybrid lattice-based model on a square lattice as an example. (b) We fit a convex hull to the final colony shape (red line). The blue dots indicate the endpoints of the convex hull segments. The normal vectors to each convex hull segment are then determined (black arrows). The weight of a normal vector (indicated by the length of the arrows) is determined by the length of its associated segment of the convex hull. This procedure is repeated for 500 independent simulations. (c) The directions and weights of all the resulting normal vectors are then collected in a histogram and normalized. Wrapping this histogram around a representative colony shape gives the figures in the main text.}
    \label{fig:convhull_procedure}
    \end{adjustwidth}
\end{figure}